\documentclass[aps,pra,twocolumn,groupedaddress,superscriptaddress]{revtex4}
 \usepackage{lipsum}
\usepackage{mathtools,amssymb,lipsum}
\usepackage{graphicx}
\usepackage{xcolor,ulem}

\newcommand{\Rfaqat}[1]{\textcolor{black}{ {\sl#1}}}

\newcommand{\be}{\begin{equation}}
\newcommand{\ee}{\end{equation}}
\newcommand{\beq}{\begin{eqnarray}}
\newcommand{\eeq}{\end{eqnarray}}
\newcommand{\bea}{\begin{array}}
\newcommand{\eea}{\end{array}}

\draft 
\begin{document}

\title{Revisit of generalized Kerker's conditions using composite metamaterials }
\author{Rfaqat Ali}
\email[]{r.ali@if.ufrj.br}
\affiliation{%
Instituto de F\'isica, Universidade Federal do Rio de Janeiro, Caixa Postal 68528, Rio de Janeiro, RJ, 21941-972, Brasil}


\begin{abstract}
Achieving zero backward scattering (ZBS) and zero forward scattering (ZFS), i.e., the so-called the first and second Kerker's conditions respectively, by sphere spherical particles is considered to be impossible due to the unavailability of naturally occurring magnetic materials in the visible frequency range. We report theoretical modeling to design composite metamaterials that present large optical magnetic permeability in the visible frequency range by employing Mie scattering theory and extended Maxwell Garnett theory.  We numerically show that a careful selection of constituents of a composite metamaterial one can obtain metamaterials with sufficiently large artificial permeability that eventually provides the Kerker's criterion to achieve the Kereker's conditions.  By taking realistic material parameters we demonstrate that the metamaterials exhibiting  ZBS and ZFS have a small imaginary part of the refractive index than metallic structures that pave a path to design high-performance nanophotonic devices.
\end{abstract}
\maketitle 

\section{Introduction}
Optimal control on directional scattering by  spherical particles in the visible frequency domain is an ultimate goal for researchers to optimize the efficiency of optical devices \cite{nano,Gao_2017,sensing,nanoen,Liu2012,plas,LED}.
 For referential examples, the desire to achieve zero backscattering  plays a crucial role in  optical manipulation of spherical particles  \cite{chan2011,kall2018,r.ali}  and light management structures used in photovoltaic devices \cite{Ferry2010,Akimov2010}.  The Mie scattering theory provides an exact solution to the  problem of light scattering by  spheres  \cite{bohren} and 
the seminal work on directional scattering was  presented by Kerker et al. in 1983 \cite{ker}.  They  revealed proper combinations of relative permittivity and relative permeability of an unconventional magneto-dielectric sphere  to achieve ZBS and ZFS that are  commonly known as the first and second Kerker's conditions, respectively.
Fulfillment of the first Kerker's condition requires a material with identical relative permittivity $\epsilon$ and relative permeability $\mu$  (i.e.  $ \epsilon=\mu)$, while the second Kerker's condition needs material with  $\epsilon= \frac{4-\mu}{\mu+2}$.
Observation of these conditions for the  visible light is considered to be impossible due to the unavailability of such materials that possess large relative magnetic permeability in the frequency range, thus, Kerker's method can not be applied in optical frequencies.

 However, in spite of these limitations ZBS and ZFS are successfully  experimentally demonstrated by  a  dipolar sphere of large permittivity   \cite{si,person}.  In this case, the incident field equally excites the electric and magnetic dipoles that produce anisotropic scattering pattern.
 Since the anisotropic distribution of the scattered field is governed by the interference between the electric and magnetic dipoles. For instance, the 
  in-phase  and out of phase oscillations  of the  electric and  magnetic dipoles can  provide an opportunity to observe  ZBS  and ZFS, respectively  \cite{Ge,Lee,neito2015,Geffrin2012}. Although these proposals  are limited to the sphere of small radius $(r)$ as compared  to the incident wavelength   $(\lambda)$, such that $r\ll \lambda$. 

Nowadays, the anomalous but very fascinating scattering behavior is observed by artificially designed so-called metamaterials with unconventional optical constants \cite{angheta2006,angheta2013}.  They  have the ability to present unique scattering properties like optical cloaking \cite{farhat}, negative optical  reflection  \cite{ap1} and  efficient control on scattering directionality \cite{r.ali} with valuable interest in the fields of nanoantenna, nano waveguides and optical manipulations \cite{r.ali,ap2,r.alee,CTchan}.  
 The recent advancement in the field of metamaterials allows to achieve the invisibility of an object through scattering cancellation by manipulating the optical permittivity and permeability of the materials \cite{farhat,soukhoulis}. Although, it is  challenging to find appropriate materials to design the metamaterial that presents a strong artificial magnetism in the visible range.
Recently, the ZBS has been shown  by a sphere made of composite metamaterial \cite{r.ali}  by tuning the permittivity of the sphere in such a way that electric and magnetic multipoles interfere destructively in the backward direction,   thus ZBS is  achieved.  This proposal  is neither  robust against the radius nor accommodates the permeability of the sphere, while the Kerker's method is applicable for all radii.

 Motivated by the recent advancement in the field of  metamaterials, in this article, we put forward a composite material platform to design  composite metamaterials that exhibit strong artificial magnetic response in the optical frequencies range. Furthermore, we engineer the optical constants  of the composite metamaterials to meet the Kerker's  criterion to achieve  the ZBS and ZFS.
 {In order to  implement this proposal, we consider a composite metamaterial made of a dielectric host medium containing dipolar spherical  inclusions of large optical permittivity \cite{farhat,Goia_2003,anghetta}. Furthermore, we  apply extended Maxwell-Garnett theory to calculate the effective optical constants of homogenized composite metamaterials  in the optical frequency range  by  assuming that the sizes of the  inclusions are sufficient to  overcome the  nonlocal effects  \cite{Ruppin1973,Gubbin_2020}. In addition, the Mie scattering theory is used to calculate the relevant directional scattering cross sections of the composite metamaterials. 
}

The rest of the article is organized as follows.  Section \ref{sec_method} is a methodological part, where the Mie scattering theory and extended Maxwell-Garnett (EMG) theory are briefly discussed. The main findings of the work are presented in sections \ref{sec_result}, where the Kerker's conditions for composite metamaterials  are analyzed. Finally, we summarize our findings  in section \ref{sec_con}.

\section{Methodology}\label{sec_method}

\subsection{Mie Theory } \label{sec_Mie}
  Consider a  plane wave with vacuum wavelength $\lambda_0$  scattering  by a spherical 
particle of radius $r$,  refractive index $ n = \sqrt{\epsilon \mu}$  embedded in a non-absorbing host medium with refractive index  $ n _1= \sqrt{\epsilon_1  \mu_1}$, where $\epsilon  (\epsilon_1)$ and 
$\mu  (\mu_1)$  are the relative permittivity and relative permeability of the sphere (host medium).
The scattering and absorption of  incident light  by the  sphere  is described by the  Mie scattering theory in terms of normalized scattering  efficiency $Q_{sca}$,   absorption efficiency  $Q_{abs}$,   extinction efficiency  $Q_{ext}$  and mathematically  defined as   \cite{bohren,Ge,Lee,van}

\begin{equation}
Q_{sca} = \frac{2}{x^2}  {\sum_{\ell=1}^\infty }{(2\ell+1)(| a_\ell|^2 +|b_\ell|^2}),
\end{equation}

  \begin{equation} 
  Q_{abs} = \frac{2}{x^2}{\sum_{\ell=1}^\infty }{(2\ell+1)( {\rm Re}{[a_\ell]}- \vert a_\ell\vert ^2 +\rm Re[b_\ell]- \vert b_\ell \vert^2}), 
  \end{equation}

\begin{equation}
Q_{ext} = Q_{abs}+Q_{sca},  \label{Q_ext}
 \end{equation} 

 where $ a_{\ell}$ and $ b_{\ell}$ are the Mie scattering coefficients corresponding  to the  transverse magnetic and
transverse electric modes, respectively. The index $\ell$ is used to denote $\ell^{th}$ order spherical harmonic channel and $x = kr$ is size parameter with $k=2\pi/\lambda$.
The Mie scattering coefficients can be derived by employing the boundary conditions on the surface of the scatterer  and  written as \cite{bohren}
\begin{eqnarray}
a_{\ell}= \frac{m\psi_{\ell}(mx) \psi'_{\ell}(x)-\mu\psi_{\ell}(x) \psi'_{\ell}(mx)}{m\psi_{\ell}(mx) \xi'_\ell(x)-\mu\xi_\ell (x) \psi'_{\ell}(mx)},  \label{Mie_Ca}
 \end{eqnarray}
 \begin{eqnarray}
  b_{\ell}=  \frac{\mu \psi_{\ell}(mx) \psi'_{\ell}(x)-m  \psi_{\ell}(x) \psi'_{\ell}(mx) }{\mu \psi_{\ell}(mx) \xi'_\ell(x)-m\xi_\ell(x)\psi'_{\ell}(mx) }, \label{Mie_Cb} 
 \end{eqnarray}
where, $\psi_\ell$, $ \xi_\ell $ are Riccati-Bessel functions \cite{Functions}, $m = n /n_1$ is the relative refractive index.
In order to study the directional scattering pattern, the expressions for differential scattering efficiencies  in  forward ($\theta=0$) and
backward ($\theta = \pi)$ directions are given as  \cite{bohren}

\begin{equation}
 Q_b|_{\theta=\pi}= \frac{1}{x^2}\bigg| { \sum_{\ell=1}^\infty }{(2\ell+1)(-1)^l(a_\ell - b_\ell)} \bigg|^2, \label{bd}
 \end{equation}
 
\begin{equation} 
Q_f|_{\theta=0}= \frac{1}{x^2}\bigg| { \sum_{\ell=1}^\infty }{(2\ell+1) (a_\ell+ b_\ell)} \bigg|^2. \label{fd}
\end{equation}

  \begin{figure*}\centering
\includegraphics[width = 5. in]{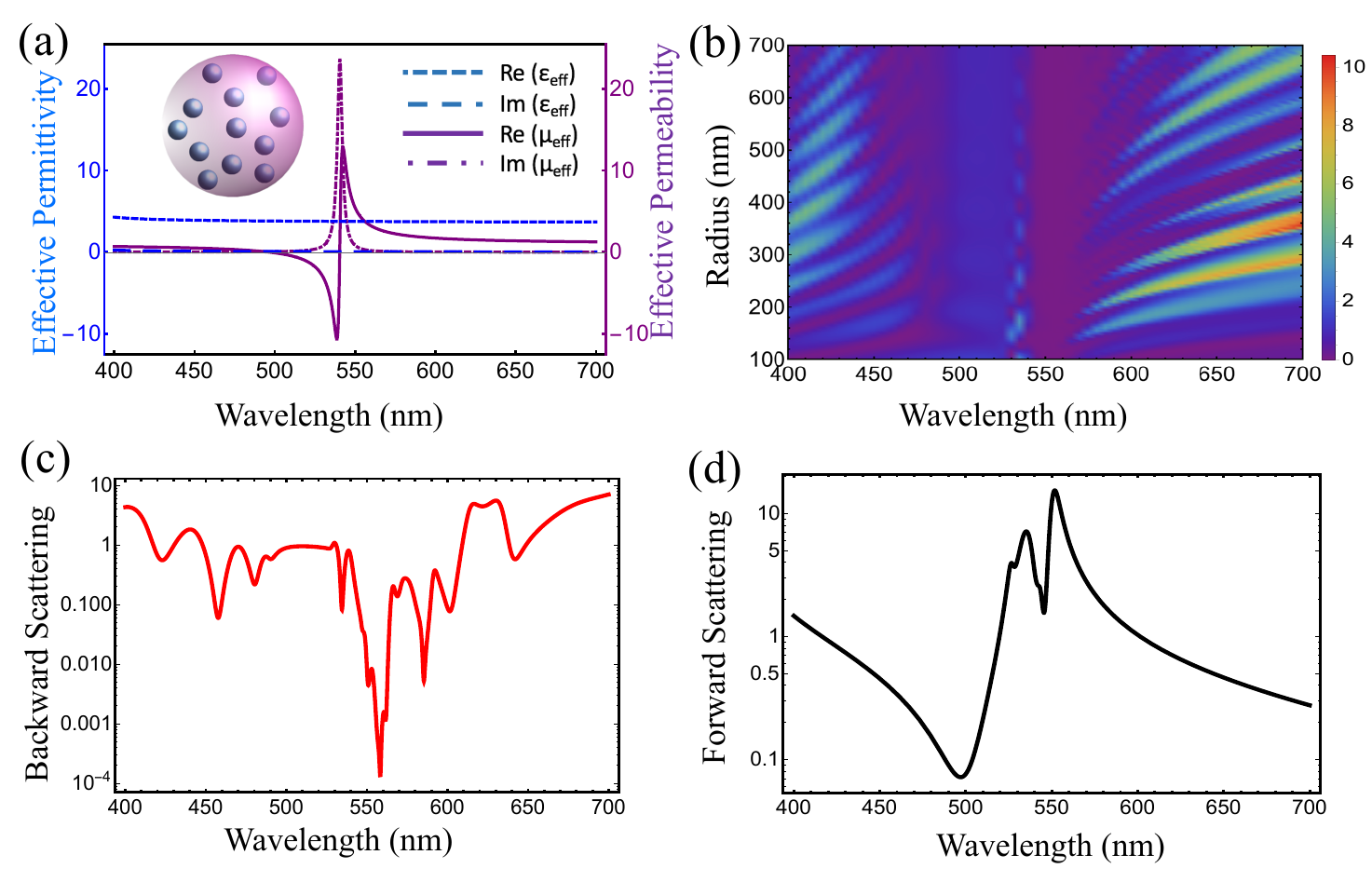} 
\caption{(a) The real and imaginary parts of the effective permittivity (blue) and  effective permeability (purple)  of the composite metamaterial  made  of  $SiO_2$  as a host medium and spherical inclusions of permittivity $\epsilon_i=169$ \cite{farhat}, where inset shows the geometry of the  composite sphere.   
(b) The  backward scattering efficiency by the composite sphere as a function of incident wavelength and radius is  plotted.  (c)
The backward scattering efficiency by  composite sphere of radius $500\, \rm nm$   and  (d) forward  scattering efficiency by composite  sphere of radius $70\, \rm nm$ are displayed as function of incident wavelength.  Throughout the figure, the radius of the spherical inclusions and volume  filling fraction are fixed at  $b=20\, \rm nm$ and  $f=0.2$ respectively.}
\label{F1}
\end{figure*}   

    It can be seen in Eq. (\ref{bd}), the first Kerker's condition (i. e. $ Q_b= 0$) can  be  achieved by providing  $a{}_\ell= b{}_\ell$ and  this can occur for identical  identical permittivity and permeability of the sphere (see Eqs. \ref{Mie_Ca}-\ref{Mie_Cb}). On the other hand, the second Kerker's  condition can be satisfied by achieving  $ Q_f\simeq0$, provided that    $a_1 \simeq -b_1$,  this occurs when the condition
    $\epsilon=\frac{4-\mu}{2\mu+1}$ is satisfied in the quasi-static limit.

{It is worthwhile to mention that}  the optical theorem imposes a {severe} condition to  achieve  ZFS. This theorem of optics relates the extinction cross section $\sigma^{ext}$ to the  forward scattering amplitude $s_\theta(0,0)$  and can be expressed  as $\sigma^{ext}= \frac{\lambda^2_0}{\pi} {\rm Im}[s_\theta(0,0)] $ \cite{NZFS,Lee_2018}.  In practice, $Q_f=0$ (at $2^{nd}$ Kerker's condition)  appears by achieving $a_1=-b_1$ without necessarily making both of them zero \cite{ker}, which not only vanishes  the forward scattering but also nullify the total extinction cross section (one may refer to Eq. (\ref{Q_ext}) in quasi-static approximation). Subsequently, the non zero  $a_1$ and $b_1$ are  implying that the overall scattering in all
other directions may still  be significantly different from zero, which seems to be a contradiction to the fact that extinction  cross section vanishes.
However, the usual  approximation for the dipole coefficients fail to satisfy  energy conservation requirements \cite{NZFS}. A more accurate approximation for those coefficients, taking into account radiative corrections,  shows that the forward scattering is not exactly zero, relaxing the contradiction. 
  Therefore, one cannot completely suppress the  forward  scattering, it  may be possible to achieve near zero forward scattering (NZFS) at second Kerker's condition.
In order  to demonstrate these conditions, we use the effective medium theory that will provide an environment to satisfy the Kerker's conditions.

 \subsection{ Extended Maxwell Garnett Theory } \label{sec_emg}
 
When electromagnetic field propagates through a heterogeneous medium comprises of non-absorbing host medium of permittivity $\epsilon_h$ and small spherical  inclusions of  permittivity $\epsilon_i$,  radius $b$, such that $b\ll \lambda$. 
 There is a variety of effective medium theories for homogenization of  the heterogeneous medium \cite{Choy2015,Doyle1989}.
The collective optical response of such a homogenized medium can be discussed by defining the effective optical constants provided by one of the effective medium theory.  
Due to the  best performance we use  extended  Maxwell-Garnett theory \cite{farhat,Muhlig_2013,ali_2020,Muhlig_2012,max2,ruppin2000}, which provides the effective permittivity and permeability as  functions of volume filling fraction ${\it f}$, size of the inclusions, permeability and permittivity of the host medium as 
   $ n_{\rm eff}= \sqrt{\epsilon_{_{\rm eff}}\mu_{_{\rm eff}}} $, where 
    
\begin{equation}
\epsilon_{_{\rm eff}}=\epsilon_{h}\frac{y^3+3if a{}_{1}}{y^3-\frac{3}{2}{i}f a{}_{1}}, \label{eeff} 
\end{equation}

 \begin{equation}
 \mu_{_{\rm eff}}=\mu_{h}\frac{y^3+3if b{}_{1}}{y^3-\frac{3}{2}if b{}_{1}}. \label{meff}  
\end{equation}
   Here  $ y= \sqrt{\epsilon}_{h} \omega b /c, $ is the size parameter inside the host medium, $\omega$ is the frequency, c is the speed of light,   $a_1 $ and  $b_1$ are dipolar Mie coefficients  of the inclusions. The size parameter  must be $y\ll 1$ \cite{ruppin2000}, otherwise { higher order multipoles }  will contribute to effective permittivity and permeability  and spoil the validity of EMG theory.

   \section{Results and Discussions } \label{sec_result}
   
We begin our numerical analysis by considering a composite material made  of  $SiO_2$   as host medium  with relative  permittivity $\epsilon_h=2.1$   \cite{sio21965} and dielectric spherical inclusions of radius $20 \,\rm nm$, bulk relative  permittivity $\epsilon_i=169$  \cite{farhat,Goia_2003} and  volume filling fraction $f$. The effective optical constants of the composite material are calculated by EMG theory (Sec. \ref{sec_emg}).  Since the inclusions have a large refractive index so that  the incident field can generate displacement currents inside the inclusions and produces tiny magnetic dipoles. The collective effect of these spherical magnetic dipoles generates a remarkable magnetic response in the composite metamaterials.
 The effective optical response of the composite metamaterials is measured in terms of effective permittivity and effective permeability by using the Eqs. (\ref{eeff}-\ref{meff}) that have  external degrees of freedom, like volume  filling fraction, permittivities and size of inclusions. One can perform fine tuning of these parameters  to tailor the resonance at a chosen wavelength for operation.
\begin{figure*}
\centering
\includegraphics[width = 4.50 in]{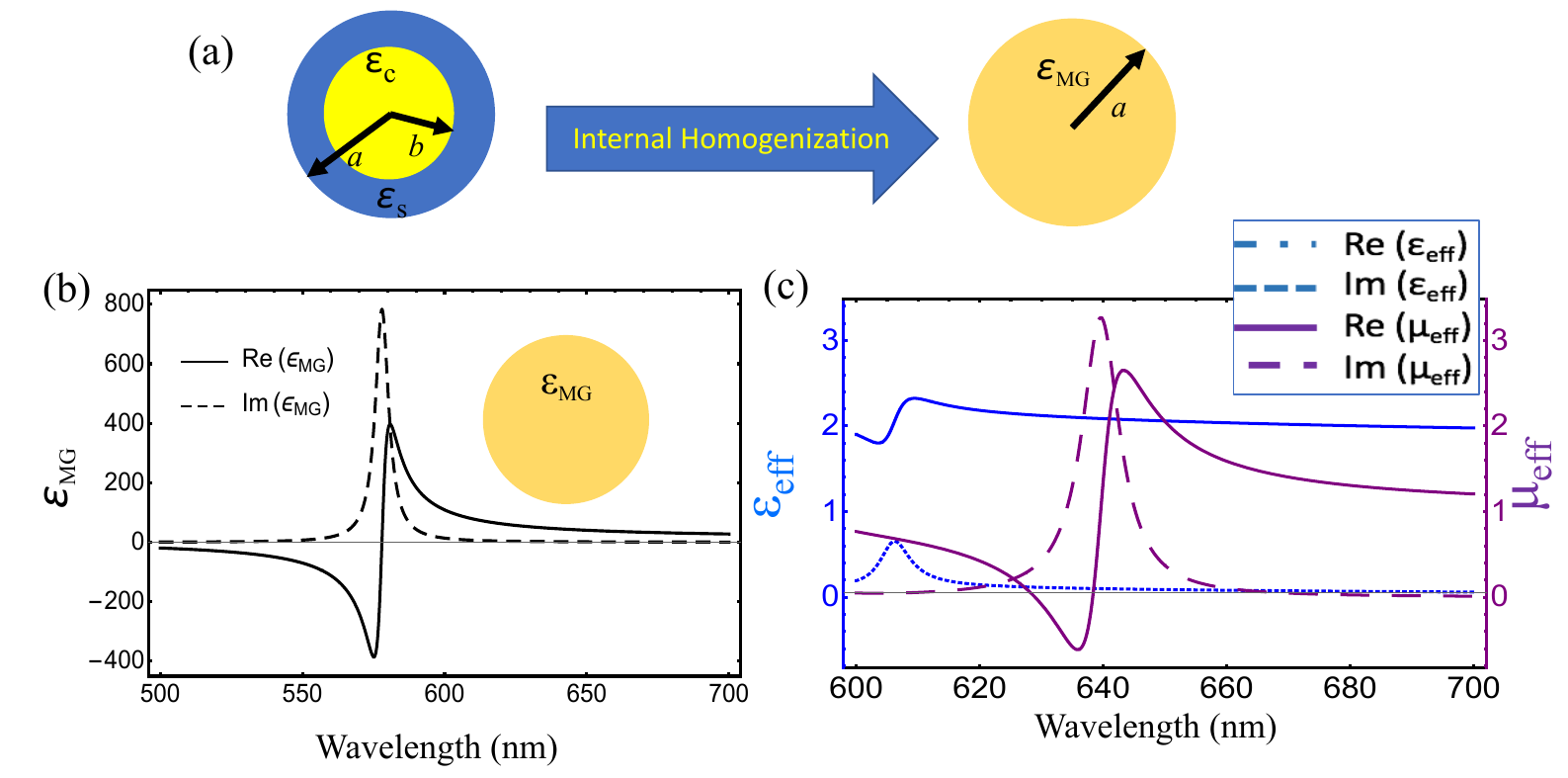}\\
\caption{(a) The schematic diagram of internal homogenization, (b) the real and imaginary parts of effective permittivity of the  effective sphere. (c)
The real and imaginary parts of effective  permittivity and permeability of an arbitrarily arranged   effective spheres in an homogeneous medium with $\epsilon_ h = \mu_h = 1$  for filling fraction $f = 0.25$ as a function of incident wavelength.  } \label{F2}
\end{figure*} 
In Fig. \ref{F1}(a), the real and imaginary parts of the effective  permittivity (blue) and permeability (purple) of the composite metamaterial are displayed as a function of  incident wavelength.
It is clearly seen that the composite metamaterial shows remarkable magnetic response with a  spectral region that indeed lies in our domain of interest. For instance,   the composite metamaterial shows an identical effective permeability and permeability at the wavelength of $560 \, {\rm nm}$ with relatively small imaginary parts.  According to Kerker's method,  this is an ideal scenario to satisfy the first Kerker's condition for all sized spherical particles.
Fig. \ref{F1}(b) shows a color map of backward scattering efficiency by the composite sphere as a function of the incident wavelength and radius of the sphere. It is clearly seen that at  $\lambda=\, 560 {\rm nm}$  the backward scattering is completely suppressed  which occurs  
 due to the  in-phase oscillations of the  electric and magnetic multipoles with  same amplitudes  such that $a_\ell=b_\ell$,  provided by  identical  $\epsilon_{\rm eff}$ and $ \mu_{\rm eff}$ as shown  in Fig. \ref{F1}(a).   Therefore, the overall backward scattering efficiency defined in Eq. (\ref{bd}) is reduced to zero regardless of the sizes of the sphere. The drastic reduction of  backward scattering by the composite sphere of radius $500\, \rm nm$ is shown in Fig. \ref{F1}(c) by calculating the backward scattering efficiency versus wavelength. It is clearly shown that  the dramatic reduction appears at  $\lambda=560\, {\rm nm}$ and backward scattering amplitude reduces to zero, thus satisfying the first Kerker's condition.

  In contrast, the second Kerker's condition requires  the condition  $\epsilon_{\rm eff}= \frac{4-\mu_{\rm eff}}{\mu_{\rm eff}+2}$,  which surprising also occurs  at a wavelength  of $500\, {\rm nm}$.  In  Fig. \ref{F1}(d), we  calculate the forward scattering efficiency by { a} composite sphere of radius  $70\, \rm nm$ as a function of the incident  wavelength. The drastic reduction of forward scattering appears at  $\lambda = 500\, {\rm nm}$ and satisfying the second Kerker's condition. It is worthwhile to  mention that the second Kerker's condition can be achieved by providing  $a_\ell= -b_\ell$, which only  implies to the dipolar  particles (i.e. for $\ell=1$).  Otherwise, this effect might be spoil for a large sphere due to higher-order multipole contributions to the  forward scattering efficiency.  Thus,   provides nearly zero forward scattering  {$Q_f\simeq0$} at  $\lambda=500\, {\rm nm}$ for a \Rfaqat{dipolar sphere (of  radius  $70\, {\rm nm})$}.

 Now we extend our proposal to design metamaterials by using naturally occurring materials that  will present a  significant magnetic response to satisfy the Kerker's conditions.
The immediate question that arises, is there any material that possesses large permittivity in the visible region? The answer is yes, for instance, a carefully designed  core-shell system shows a very high permittivity due to plasmonic resonances  with strong spectral dependence and in the quasi-static approximation these core-shell nanoparticles  can be used as inclusions into a host medium.
 For simplicity,  
we begin with an internal homogenization approach introduced by Chettiar and Enghetta \cite{anghetta} that provides effective permittivity of core-shell nanoparticle by homogenizing it  to an effective sphere of the same dimension.    Through this approach, the effective permittivity of a core-shell nanoparticle is calculated by equating  the polarizabilities  of the core-shell   to  a homogenized sphere of the  same dimension and finally, the expression  turns out as 
\begin{equation}
\epsilon_{\rm MG}= \epsilon_{s}\frac{a^3(\epsilon_{c}+2\epsilon_{s})+2b^3(\epsilon_{c}-\epsilon_{s})}{a^3(\epsilon_{c}+2\epsilon_{s})-b^3(\epsilon_{c}-\epsilon_{s})}. \label{mgc} 
\end{equation} 

\begin{figure*}
\centering
\includegraphics[width = 4.5 in]{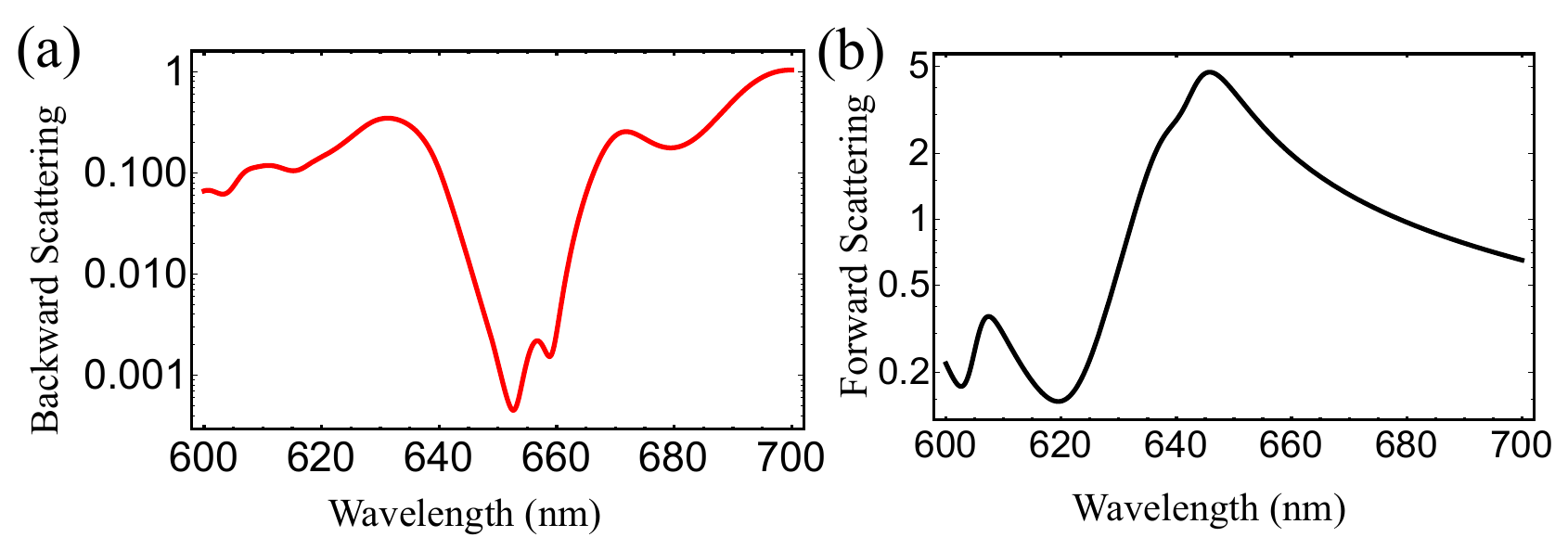}\\
\caption{(a) Backward scattering efficiency by a sphere made of composite  metamaterial  (of permittivity given in Fig. \ref{F2}c) of radius $800 \, {\rm nm}$ as a function of incident wavelength.   (b) Forward scattering efficiency by a composite sphere of radius $120 \, {\rm nm}$, where the volume filling fraction is fixed at $f=0.25.$}
\label{F3} 
\end{figure*}   

\Rfaqat{ In order to implement this proposal, we consider  metallic  nanosphere  as  a core of radius $b$  embedded in  dielectric  spherical shell of radius $a,$   forming  a core-shell nanoparticle  \cite{Haguraa_2010,Oldenburg_1992,Chaudhuri_2012,Hanske}. Assuming that the core  is silver ($\rm Ag$) and we may use  Drude formula to define it's  permittivity,  $\epsilon_c=\epsilon_\infty-\frac{\omega^2_p}{\omega(\omega+i\Gamma)}$, with  $\omega_p = 9.2 eV,$  $ \Gamma = 0.0212\, eV,$ $ \epsilon_\infty = 5.0,$ where $1 eV = 241.8\, THz $  \cite{Johnson_1972} and $\omega$ is the angular frequency of incident light. Furthermore,} the core-shell is  homogenized to an effective sphere of radius $a$ with  permittivity $\epsilon_{\rm MG}$ provided by Eq. (\ref{mgc}) and displayed   in Fig. \ref{F2}(b) at fixed radii ratio at $\frac{b}{a} =0.84$. It has been verified that in the quasi-static limit,  the effective sphere has  a good agreement to exact  Mie solution of a core-shell nanoparticles of the same dimension \cite{crystal} and indeed it can be used as inclusion into a host medium to fabricate the structures of desired optical and physical properties \cite{anghetta,crystal,JJS,Rapid}.

Let \Rfaqat{us} design a bulk composite medium comprises of randomly arranged effective (homogenized) nanospheres  with
permittivity $\epsilon_{\rm MG}$ and radius $a$  embedded in a host medium with volume filling fraction $f$. 
In Fig.  \ref{F2}(c) we calculate  the   effective permittivity (blue) and permeability (purple)  of  newly designed composite metamaterial  by means of
EMG theory at fixed $ f=0.25$ as a function of $\lambda$. In this configuration, we consider radii of the effective nanosphere about 14 time less than the incident wavelength which is a safe zone for the validation of the effective medium theory \cite{ruppin2000}. 

It can be seen in Fig. \ref{F2}(c)  that the composite metamaterial not only presents  large  effective permeability but also \Rfaqat{provides}  identical permittivity and permeability with  week imaginary parts at a wavelength of $650\, {\rm nm}$.
Therefore, according to Kerker's criterion, it must present zero backward  scattering at this wavelength.
We now calculate the directional scattering by the composite sphere of  radius $r$  with permittivity given  in  Fig. \ref{F2}(c).   The  backward scattering efficiency and forward scattering efficiency as a function of incident wavelength are displayed  in Fig. \ref{F3}(a) and \ref{F3}(b), respectively.  It is clearly seen that the backward scattering efficiency is drastically reduced to zero at $\lambda=650 \, {\rm nm}$ which is due to the fact that equal electric and magnetic response. On the other hand,  \Rfaqat{the NZFS appears  at a  wavelength of  $620\, {\rm nm}$ due to out of phase oscillations of electric and  magnetic dipoles, thus satisfying the second Kerker's condition.}

\section{Conclusions} \label{sec_con}
In this paper, we have theoretically modeled composite metamaterials using  Mie scattering theory and effective medium theory. Our numerical results have shown that the composite  metamaterials present strong artificial magnetic permeability in the optical frequency domain that appears due to the displacement currents inside the inclusions.
Our findings show that  a fine tuning of the parameters appear in  the EMG theory,  one can creates  a  spectral region with identical permittivity and permeability not only for bulk homogeneous  composite  spherical sphere  but also for a regularly arranged \Rfaqat{nano}spherical particles. \Rfaqat{Altogether, this proposal not only paves a path to design new possible structures with artificial magnetic response but also  allows to achieve Kerker's conditions for low refractive index materials in the visible frequency range.}   
These structures with unique optical properties have very low ${\rm  Im( m_{eff})}$ (i. e. low absorption) than metallic \Rfaqat{ structure} that make\Rfaqat{s} our proposal a good candidate to design high-performance optical antennas,  metamaterials and other novel nanophotonic devices.
\section{ Acknowledgment }
We thank F. Pinheiro, F. Rosa, PAM Neto and 
S. Iqbal for inspiring discussions. This work has
been supported by the Brazilian agencies National Council for Scientific and Technological Development (CNPq),
Coordination for the Improvement of Higher Education
Personnel (CAPES).

\end{document}